\documentclass[aps,prb,reprint]{revtex4-1}
\usepackage{bm,amssymb}
\usepackage{amsmath}    % need for subequations
\usepackage{graphicx}   % need for figures
\newcommand{\simg}{\stackrel{>}{_\sim}}
\newcommand{\siml}{\stackrel{<}{_\sim}}
\begin{document}
\title{
Dynamical mean-field theory for the anisotropic Kondo semiconductor:\\ 
Temperature and magnetic field dependence
%Finite temperature and magnetic field effects on the anisotropic Kondo semiconductor by using the dynamical mean-field theory with exact diagonalization
}
%\author{Takemi Yamada and Yoshiaki \=Ono}
\author{Takemi Yamada}\email{takemi@phys.sc.niigata-u.ac.jp}
\author{Yoshiaki \=Ono}
%\thanks{takemi@phys.sc.niigata-u.ac.jp}
\affiliation{
Department of Physics, Niigata University, Ikarashi, Nishi-ku, Niigata, 950-2181, Japan
}
\date{\today}

\begin{abstract}
We investigate the periodic Anderson model with $\bm{k}$-dependent $c$-$f$ mixing reproducing the point nodes of the hybridization gap 
by using the dynamical mean-field theory combined with the exact diagonalization method. 
At low temperature below a coherence temperature $T_0$, the imaginary part of the self-energy is found to be proportional to $T^2$ 
and the pseudogap with two characteristic energies $\tilde{\it \Delta}_1$ and $\tilde{\it \Delta}_2$ is clearly observed for $T\ll T_0$, 
while the pseudogap is smeared with increasing $T$ and then disappears at high temperature $T \simg T_0$ due to the evolution of the imaginary self-energy. 
When the Coulomb interaction between $f$ electrons $U$ increases, $\tilde{\it \Delta}_1$, $\tilde{\it \Delta}_2$, and $T_0$ together with $T_{\rm max}$ 
at which the magnetic susceptibility is maximum decrease in proportion to the renormalization factor $Z$ resulting in a heavy-fermion semiconductor 
with a large mass enhancement $m^*/m=Z^{-1}$ for large $U$. 
We also examine the effect of the external magnetic field $H$ 
and find that the magnetization $M$ shows two metamagnetic anomalies $H_1$ and $H_2$ corresponding to $\tilde{\it \Delta}_1$ and $\tilde{\it \Delta}_2$ 
which are reduced due to the effect of $H$ together with $Z$. Remarkably, $Z^{-1}$ is found to be largely enhanced due to $H$ especially for $H_1 \siml H \siml H_2$, 
where the field induced heavy-fermion state is realized. 
The obtained results seem to be consistent with the experimental results observed in the anisotropic Kondo semiconductors such as CeNiSn. 
\end{abstract}
%\keywords{}

\maketitle
\section{Introduction\label{sec:1}}%%%%%%%%%%%%%%%%%%%%%%%%%%%%%%%%%%%%%%%%%%%%%%%%%%
In the heavy-fermion systems, $f$ electrons hybridize with conduction $(c)$ electrons via the $c$-$f$ mixing to form coherent quasiparticles with large effective mass, 
which is due to the effect of Coulomb interaction between $f$ electrons at low temperature, 
while at high temperature, $f$ electrons are almost localized and scatter $c$ electrons resulting in the Kondo effect.\cite{rev-HF} 
The systems show various types of ground states including the so-called Kondo semiconductor, 
which exhibits an insulating behavior at low temperature with highly reduced energy gap. 
With increasing temperature, the energy gap tends to be smeared, and then the system shows the behavior of incoherent metal at high temperature. 
Typical examples of the Kondo semiconductors are SmB$_6$,\cite{SmB6} 
YbB$_{12}\cite{YbB12}$ and Ce$_3$Bi$_4$Pt$_3$,\cite{Ce3Bi4Pt3} which 
have cubic crystal structures and possess well defined energy gaps of the orders of 100 K as observed in the measurements of the thermodynamic and transport properties. 

%Another class of the Kondo semiconductors such as CeNiSn and its isostructural compounds with the orthorhombic structure shows the behavior of anisotropic semiconductor,\cite{CeNiSn1,CeNiSn2,CeNiSn3,CeRhSb,CeRhAs1} 
%where the energy gap vanishes in specific regions of the $\bm{k}$-space, and then a pseudogap is observed in the electronic density of states (DOS) at low temperature below $\sim$10 K. 
%Experimentally, such a pseudogap was first observed from the measurements of the longitudinal NMR relaxation rate\cite{NMR1,NMR2,NMR3} at low temperature and supported later by those of the thermodynamic properties.\cite{CeNiSn3,C/T1,C/T2} 
%In addition, the inelastic neutron scatterings show the existence of anisotropic magnetic excitations in the $\bm{k}$-space.\cite{NSCAT1,NSCAT2,NSCAT3} 
%The anisotropic pseudogap behavior has also been observed in the magnetization and magnetoresistance.\cite{MAG1,MAG2,MAG3} 
%The pseudogap structure can be directly observed by the tunneling\cite{TS} and photoemission\cite{CeRhAs2,CeRhAs3} spectroscopies which exhibit significant temperature dependence of the gap structure, where the gap is clearly observed at low temperature while it vanishes at high temperature. 
%More recently, the pressure\cite{CeNiSn-P,CeRhAs-P} and doping\cite{DOPED1,DOPED2,DOPED3} effects of the pseudogap have been extensively investigated together with the possible long range order.\cite{CeRhAs1} 

Another class of the Kondo semiconductors such as CeNiSn and its isostructural compounds with the orthorhombic structure shows the behavior of anisotropic semiconductor 
or semimetal.\cite{CeNiSn1,CeNiSn2,CeNiSn3,CeRhSb,CeRhAs1,NMR1,NMR2,NMR3,C/T1,C/T2} 
The longitudinal NMR relaxation rate $1/T_{1}$\cite{NMR1,NMR2,NMR3} and the Sommerfeld coefficient $\gamma$\cite{CeNiSn3,C/T1,C/T2} are suppressed below 10 K 
indicating the development of a pseudogap in the density of states (DOS) at low temperature. 
Such pseudogap behavior is well accounted for by the V-shaped gap model with a residual DOS\cite{NMR1} or the semimetallic model with nodes in the gap.\cite{Ikeda1996} 
The V-shaped pseudogap in the DOS was directly observed below 10 K in the tunneling spectroscopy.\cite{TS} 
As for the transport properties, 
the resistivity along the $a$-axis $\rho_{a}$ decreases with decreasing temperature as expected by the semimetallic model,\cite{Ikeda1996} 
while $\rho_{b}$ and $\rho_{c}$ slightly increase below 3 K.\cite{CeNiSn3} 
The inelastic neutron scattering experiments revealed the existence of anisotropic magnetic excitations.\cite{NSCAT1,NSCAT2,NSCAT3} 
Anisotropic pseudogap properties were also observed in the magnetization and the magnetoresistance.\cite{MAG1,MAG2,MAG3} Despite the intense efforts, 
it is still controversial whether CeNiSn has a zero DOS just at the Fermi level or a dip structure DOS around the Fermi level within the experimental uncertainties. 

A remarkable feature of the anisotropic Kondo semiconductors is the significant temperature dependence of the pseudogap. 
In the tunneling\cite{TS} and the photoemission spectroscopies,\cite{CeRhAs2,CeRhAs3} the pseudogap in the DOS is clearly observed at low temperature, 
while it vanishes at high temperature. Such temperature dependent pseudogap cannot be explained with a simple rigid-band model. 
Therefore the electron correlation effect is considered to be crucial for the temperature dependence of the pseudogap together with the large reduction of the gap width. 
More recently, possible long-range ordered states in the anisotropic Kondo semiconductors have also been extensively investigated 
with the effects of the pressure\cite{CeNiSn-P,CeRhAs-P} and the doping.\cite{DOPED1,DOPED2,DOPED3} 

Many theoretical studies for the Kondo semiconductor have been made on 
the basis of the periodic Anderson model (PAM)\cite{PAM1,PAM2,PAM3} with 
$\bm{k}$-independent $c$-$f$ mixing reproducing the isotropic 
hybridization gap by means of various methods such as the Gutzwiller 
approximation,\cite{GUT1,GUT2} the slave-boson mean-field 
theory,\cite{SB1,SB2,SB3} the noncrossing approximation,\cite{NCA} the 
$1/N$  expansion\cite{1/N0,1/N1,1/N2} and the dynamical mean-field theory (DMFT).\cite{DMFT-rev,DMFT1,DMFT2,DMFT3,DMFT4,DMFT5} 
These studies have shown that, due to the strong correlation effect, 
the hybridization gap is highly reduced to form a renormalized gap\cite{GUT1,GUT2,SB1,SB2,SB3} which is clearly observed at low temperature 
but disappears at high temperature.\cite{NCA,1/N0,1/N1,1/N2,DMFT-rev,DMFT1,DMFT2,DMFT3} 
A magnetic field induced insulator to metal transition has also been observed at low temperature.\cite{DMFT4,DMFT5} 
The obtained results are consistent with the experimental results observed in the isotropic Kondo semiconductors such as Ce$_3$Bi$_4$Pt$_3$ and YbB$_{12}$. 

As for the anisotropic Kondo semiconductors such as CeNiSn, 
the $\bm{k}$-dependence of the $c$-$f$ mixing is considered to be important in addition to the strong correlation effect. 
The $\bm{k}$-dependent $c$-$f$ mixing originates from the crystal electric field (CEF) ground states of $f$ electrons\cite{PAM2,PAM3,CF-MIX} 
and yields the specific DOS with the pseudogap structure.\cite{Ikeda1996,Moreno2000,Hanzawa2002,Miyazawa2003}
The PAM with the $\bm{k}$-dependent $c$-$f$ mixing has been studied by using the Gutzwiller approximation\cite{Ikeda1996} 
and the slave-boson mean-field theory,\cite{Moreno2000} 
which reproduce the highly reduced pseudogap and well explain the thermodynamic and transport properties of the anisotropic Kondo semiconductors at low temperature. 
However, the temperature dependence of the pseudogap, which is directly observed in the tunneling and photoemission spectroscopies,\cite{TS,CeRhAs2,CeRhAs3} 
together with the magnetic field dependence was not discussed there. 

The purpose of this paper is to elucidate the effects of temperature and magnetic field on the electronic states of the anisotropic Kondo semiconductors. 
For this purpose, we study the PAM with $\bm{k}$-dependent $c$-$f$ mixing by using the DMFT which becomes exact in the limit of infinite spatial dimensions and is expected to be a good approximation in three dimensions. 
The DMFT is known to describe well the strongly correlated electron systems over the whole parameter regime of temperature, magnetic field and frequency, and has been extensively developed for the PAM with $\bm{k}$-independent $c$-$f$ mixing to describe the heavy-fermion systems and the Kondo semiconductors.\cite{DMFT-rev,DMFT1,DMFT2,DMFT3,DMFT4,DMFT5} 
In the previous work, we have employed the DMFT combined with the exact diagonalization (ED) method for the PAM with the $\bm{k}$-dependent $c$-$f$ mixing and have obtained the magnetic field dependence of the electronic state which well accounts for the metamagnetic behavior observed in CeRu$_2$Si$_2$.\cite{TY1} 
The present paper is a straight forward extension of the previous work for the case with the anisotropic Kondo semiconductors such as CeNiSn. 
%Here, we adopt the same assumption of Ikeda and Miyake \cite{Ikeda1996} for the model of CeNiSn, though it remains unknown that whether CeNiSn has a zero in the DOS or as a dip characteristic of a semimetal. 

In this paper, we investigate the anisotropic Kondo semiconductor on the basis of the PAM with the $\bm{k}$-dependent $c$-$f$ mixing at half-filling 
by using the DMFT+ED method.\cite{TY1} 
The physical quantities are calculated systematically over the wide parameter regime of temperature $T$, magnetic field $H$ 
and Coulomb interaction $U$ between $f$ electrons. 
The paper is organized as follows: in Sec. \ref{sec:2}, 
we present the Hamiltonian of the PAM with the $\bm{k}$-dependent $c$-$f$ mixing and the formulation of the DMFT+ED method. 
In Sec. \ref{sec:3}, we show the results of the physical quantities for $H=0$, the renormalized DOS, the magnetic and charge susceptibilities, 
the renormalization factor and the imaginary part of the self-energy as functions of $U$ and $T$. 
In Sec. \ref{sec:4}, we present the $H$ dependence of the physical quantities, the magnetization and the renormalization factor for various $U$ and $T$. 
In Sec. \ref{sec:5}, the paper is ended with a summary together with discussions 
where the present results are compared with the previous theoretical results and the experimental results.

\section{Model and Formulation\label{sec:2}}%%%%%%%%%%%%%%%%%%%%%%%%%%%%%%%%%%%%%%%%%
\subsection{Model Hamiltonian\label{subsec:2A}}
Our model Hamiltonian of the PAM with $\bm{k}$-dependent $c$-$f$ mixing\cite{CF-MIX,Ikeda1996,Miyazawa2003} consists of the conduction electron term $H_{c}$, 
the $f$ electron term $H_{f}$ and the $c$-$f$ mixing term $H_{cf}$ as follows
\begin{align}
H&=H_{c}+H_{f}+H_{cf}, \label{eq:H}\\
&H_{c}  = \sum_{\bm{k}\sigma}\epsilon_{\bm{k}}c_{\bm{k}\sigma}^{\dagger}c_{\bm{k}\sigma},\label{eq:Hc} \\ 
&H_{f}  = \sum_{im}\epsilon_{fm}n^{f}_{im}+U\sum_{i}n^{f}_{i+}n^{f}_{i-},\label{eq:Hf} \\
&H_{cf} = \sum_{\bm{k}m\sigma}\left(V_{\bm{k}m\sigma} f_{\bm{k}m}^{\dagger}c_{\bm{k}\sigma}+{\rm H.c.}\right),\label{eq:Hcf}
\end{align}
where $c_{\bm{k}m}^{\dagger}$ is a creation operator for a $c$ electron with the wave vector $\bm{k}$ 
and the spin $\sigma=\uparrow,\downarrow$, $f_{im}^{\dagger}$ is that for a $f$ electron with the lowest Kramers doublet state $m=\pm$ at site $i$ 
and $n^{f}_{im}=f_{im}^{\dagger}f_{im}$. 
$\epsilon_{\bm{k}}$ ($\epsilon_{fm}$) is the energy for the $c$ ($f$) electron, $V_{\bm{k}m\sigma}$ is the $c$-$f$ mixing matrix element 
and $U$ is the Coulomb interaction between $f$ electrons. In Eq. (\ref{eq:Hcf}), 
the effect of the external magnetic field $H$ is included only in the $f$ electrons as $\epsilon_{fm}=\epsilon_{f}-m H$, 
because the $g$ value for the $f$ electron is known to be much larger than that for the $c$ electron. 

In this paper, we assume that the lowest Kramers doublet state under the CEF is $J_z=\pm3/2$, 
which is referred to as $m=\pm$ and the $c$ electron state is simply given by the plane wave. 
Such a model was originally developed by Ikeda and Miyake to describe the electronic state of the anisotropic Kondo semiconductors such as CeNiSn.\cite{Ikeda1996}

In this case, the $c$-$f$ mixing matrix element is given by [see Eq. (\ref{eq:Ik0}) in Appendix] 
\begin{eqnarray}
\left(\begin{array}{cc}
V_{\bm{k}+\uparrow} &V_{\bm{k}+\downarrow}  \\
V_{\bm{k}-\uparrow} &V_{\bm{k}-\downarrow}  \\
\end{array}\right)
=V_{cf}\left(\begin{array}{cc}
-a Y_{31}(\Omega_{\bm{k}}) & b Y_{32}(\Omega_{\bm{k}}) \\
-b Y_{3-2}(\Omega_{\bm{k}}) & a Y_{3-1}(\Omega_{\bm{k}})
\label{eq:Vkm}
\end{array}\right)
\end{eqnarray}
with $a=\sqrt{\frac{8\pi}{7}}$ and $b=\sqrt{\frac{20\pi}{7}}$, where $Y_{pq}(\Omega_{\bm{k}})$ is a spherical harmonics with the argument of the
solid angle $\Omega_{\bm{k}}$ of the wave vector $\bm{k}$, and $V_{cf}$ is the $c$-$f$ mixing strength defined in Eq. (\ref{eq:Vcf}) in Appendix, 
which is a parameter in our model.

\subsection{$c$-$f$ hybridized bands for $U=0$\label{subsec:2B}}
\begin{figure}[t]
\begin{center}
\includegraphics[width=7.00cm]{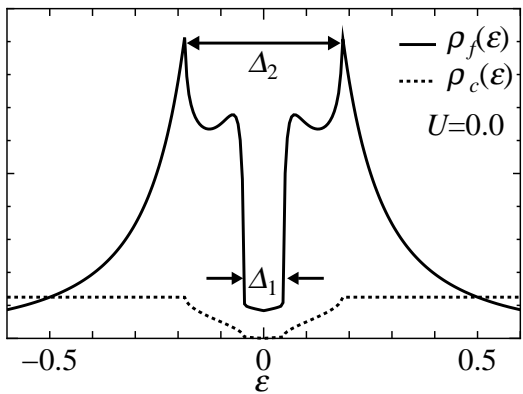}
\vspace{-0.5cm}
\caption{The noninteracting DOS for $f$ and $c$ electrons $\rho_f(\varepsilon)$ 
and $\rho_c(\varepsilon)$ near $\varepsilon=0$ for $U=0$, $\epsilon_{fm}=0$, $V_{cf}=0.5$ and $D=2$.}\label{fig:b-dos}
%\vspace{+0.2cm}
\end{center}
\end{figure}

In the noninteracting case with $U=0$, 
the Hamiltonian Eqs. (\ref{eq:H})$-$(\ref{eq:Hcf}) with Eq. (\ref{eq:Vkm}) is diagonalized to yield the $c$-$f$ hybridized bands with the energies, 
\begin{align}
E_{\bm{k}m}^{(\pm)}=\frac{1}{2}\left(\epsilon_{fm}+\epsilon_{\bm{k}}\pm\sqrt{(\epsilon_{fm}-\epsilon_{\bm{k}})^{2}+4I_{\bm{k}}}\right) 
\end{align}
with $I_{\bm{k}}=\sum_{\sigma}|V_{{\bm k}m\sigma}|^{2}$, 
where $I_{\bm{k}}$ depends only on the $z$ component of the unit $\bm{k}$-vector, $\hat{k}_z=k_{z}/|\bm{k}|$, 
and is explicitly given by (see Eq. (\ref{eq:Ik}) in the Appendix)
\begin{align}
I_{\bm{k}}=\frac{3}{8}V_{cf}^{2}(1-\hat{k}_{z}^{2})(1+15\hat{k}_{z}^{2})\label{eq:Vk}.
\end{align}
Then, the hybridization gap between the upper and the lower hybridized 
bands has nodes on the $k_z$ axis with $\hat{k}_{z}=\pm 1$ resulting in 
a pseudogap structure of the DOS as shown in Fig. \ref{fig:b-dos}, 
where two characteristic energies of the pseudogap ${\it \Delta}_1$ and ${\it \Delta}_2$ originates from the minimum of $I_{\bm{k}}$ at $\hat{k}_z=0$ 
and the maximum of $I_{\bm{k}}$ at $\hat{k}_z=\pm \sqrt{\frac{7}{15}}$, respectively. 
This pseudogap is found to well reproduce the anisotropic Kondo semiconductor CeNiSn,\cite{Ikeda1996} 
where the characteristic temperature dependence of the specific heat and the NMR relaxation rate at low temperature is well accounted for by the pseudogap. 
We note that the $f$ electrons DOS at the Fermi level is finite as shown in Fig.$~$\ref{fig:b-dos}, 
and then, the resistivity shows a metallic behavior at low temperature as observed in CeNiSn.\cite{Ikeda1996} 
This is a striking contrast to the case with the $\bm{k}$-independent $c$-$f$ mixing 
that yields a finite hybridization gap reproducing the isotropic Kondo semiconductor, 
where the resistivity shows a semiconducting behavior at low temperature. 
Here and hereafter, we assume the bare $c$-DOS to be a rectangular DOS with the band width 2$D$ centered at $\varepsilon=0$.

%Fig.1%%%%%%%%%%%%%%%%%%%%%%%%%%%%%%%%%%%%%%%%%%%%%%%%%%%%%%%%%%%%%%%%%%%%%%%%%%%%%%%
%%%%%%%%%%%%%%%%%%%%%%%%%%%%%%%%%%%%%%%%%%%%%%%%%%%%%%%%%%%%%%%%%%%%%%%%%%%%%%%%%%%%%

\subsection{DMFT+ED formalism\label{subsec:2C}}
In the DMFT,\cite{DMFT3,DMFT4,DMFT5,TY1} 
the lattice model is mapped onto an effective impurity model embedded in an effective medium which is to be determined self-consistently. 
To solve the effective impurity model, we employ the ED method for a finite-size cluster given by the following Hamiltonian, 
\begin{align}
H_{\rm imp}=&\sum_{m}\epsilon_{0m}n_{0m}+U n_{0+}n_{0-}+\sum_{l=1}^{N_s-1}\sum_{m}\epsilon_{lm}n_{lm}\nonumber\\
&+\sum_{l=1}^{N_s-1}\sum_{m}\left(V_{lm}a^{\dagger}_{lm}a_{l-1m}+{\rm H.c.}\right),\label{eq:EIM}
\end{align}
where $a_{lm}^{\dagger}$ is a creation operator for an electron with $m=\pm$ for the impurity site $l=0$ 
and that for the effective medium sites $l=1,\cdots,N_s-1$, respectively, and $n_{lm}=a_{lm}^{\dagger}a_{lm}$. 
$U$ is the Coulomb interaction between electrons on the impurity site and is set to be the same value of $U$ in the original lattice Hamiltonian Eq. (\ref{eq:Hf}). 
A set of parameters $\verb|{| \epsilon_{lm},~V_{lm} \verb|}|$ is so-called Weiss field parameters (WFPs), 
which represents the effective medium and is to be determined self-consistently. 

In the noninteracting case with $U=0$, the impurity Green's function is written with the WFPs as
\begin{align}
{\cal G}_{m}^{0}(z_{\nu})=\left(z_{\nu}-\epsilon_{0m}-\sum_{l=1}^{N_{s}-1}\frac{|V_{lm}|^{2}}{z_{\nu}-\epsilon_{lm}}\right)^{-1},
\label{eq:WFP}
\end{align}
where $z_{\nu}=i(2\nu+1)\pi T$ is the Matsubara frequency with the temperature $T$. 
For finite $U$, we solve the $N_s$-site Hamiltonian (\ref{eq:EIM}) by using the Householder ED algorithm to obtain the impurity Green's function
$G_m(z_{\nu})=({\cal G}_{m}^{0}(z_{\nu})^{-1}-\Sigma_{m}(z_{\nu}))^{-1}$ together with the impurity self-energy $\Sigma_{m}(z_{\nu})$. 
Then, the self-consistency condition, 
where the impurity Green's function coincides with the local $f$ electron Green's function $G_{m}^{f}(z_\nu)$ of the original PAM in Eq. (\ref{eq:H}) 
with the same self-energy $\Sigma_{m}(z_{\nu})$, is given by
\begin{align}
G_{m}^{f}(z_\nu)&=\frac{1}{N}\sum_{\bm{k}}\left(z_{\nu}-\epsilon_{fm}-\Sigma_{m}(z_{\nu})-\frac{I_{\bm{k}}}{z_{\nu}-\epsilon_{{\bm k}}}\right)^{-1} \nonumber\\
   &=\frac{1}{{\cal G}_{m}^{0}(z_{\nu})^{-1}-\Sigma_{m}(z_{\nu})}. 
\label{eq:sce}
\end{align}
Substituting Eq. (\ref{eq:Vk}) into $G_{m}^{f}(z_\nu)$ in Eq. (\ref{eq:sce}), we obtain a more explicit expression for $G_{m}^{f}(z)$ as
\begin{align}
G_{m}^{f}(z)&=\frac{1}{z-\epsilon_{fm}-\Sigma_{m}(z)}+\frac{1}{(z-\epsilon_{fm}-\Sigma_{m}(z))^2}\nonumber\\
&~\times\frac{1}{2D}\int_{0}^{1}d\hat{k}_{z}~I_{\bm{k}}{\rm ln}\left(\frac{\zeta+D}{\zeta-D}\right) \label{eq:gf}
\end{align}
with
$\zeta= z_\nu-I_{\bm{k}}/(z_\nu-\epsilon_{fm}-\Sigma_{m}(z))$.

In the explicit calculation to obtain the DMFT+ED solution, the following procedures are carried out: 
(i) under given WFPs $\verb|{| \epsilon_{lm}^{\rm old},~V_{lm}^{\rm old} \verb|}|$, 
the $N_s$-site Hamiltonian Eq. (\ref{eq:EIM}) is solved by using the Householder ED algorithm to obtain eigen values and eigen vectors.  
(ii) From the eigen values and eigen vectors, the self-energy $\Sigma_{m}(z_{\nu})$ is calculated. 
(iii) Substituting $\Sigma_{m}(z_{\nu})$ into Eq. (\ref{eq:sce}), 
new WFPs $\verb|{| \epsilon_{lm}^{\rm new},~V_{lm}^{\rm new} \verb|}|$ are determined 
so as to satisfy the self-consistency condition Eq. (\ref{eq:sce}) with Eq. (\ref{eq:WFP}) as possible. 
The steps (i)$-$(iii) are iterated until the old and new WFPs coincide with each other. 

\section{Results for $H=0$ \label{sec:3}}%%%%%%%%%%%%%%%%%%%%%%%%%%%%%%%%%%%%%%%%%%%%%%%%%%%%%%%
In this section, we show the $U$ and $T$ dependence of physical quantities in the absence of the external magnetic field, $H=0$. 
We restrict ourselves only to the half-filling case with the particle-hole symmetry, 
where we set $\epsilon_{f}=-U/2$ and the $c$ and $f$ electron numbers per site are given by $\langle n^c\rangle=\langle n^f\rangle=1$. 
In addition, we assume that the system is in the paramagnetic state with $\langle n_{+}^{f}\rangle=\langle n_{-}^{f}\rangle=\frac{1}{2}$, 
and then the physical quantities are independent of $m$. 
Here and hereafter, the parameters are set to as follows: the half $c$-band width $D=2$, 
the $c$-$f$ mixing strength $V_{cf}=0.5$, and the site number of the effective impurity model $N_s=6$.
\footnote[0]{
We have confirmed that the obtained results are almost unchanged for $N_s=5$ and 6 at finite temperature, and also for $N_s=6$ and 8 at zero temperature.
}

\subsection{Renormalized $f$-DOS\label{subsec:3A}}
\begin{figure}[t]
\begin{center}
\includegraphics[width=8.5cm]{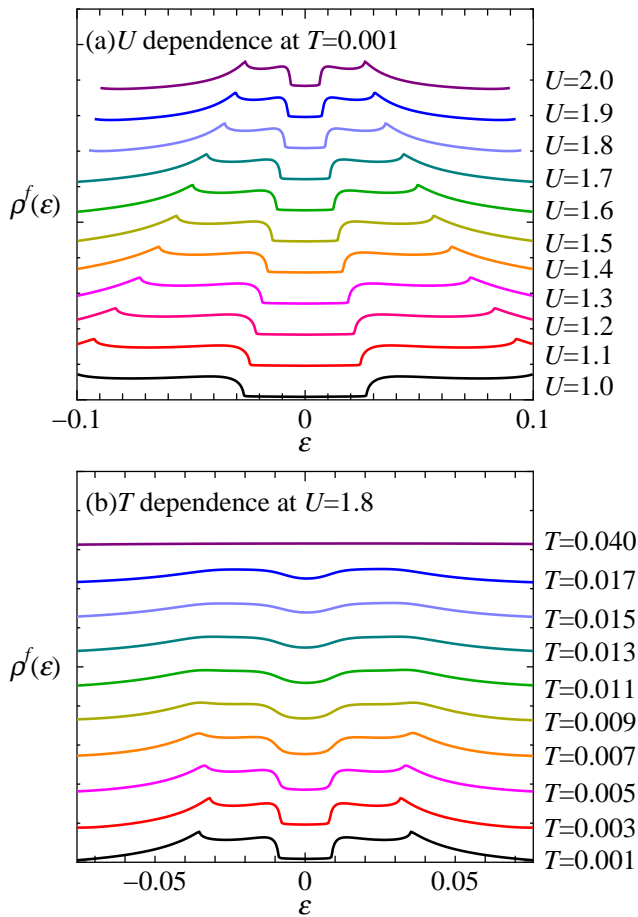}
\vspace{-0.5cm}
\caption{(Color online) The renormalized $f$-DOS $\rho^{f}(\epsilon)$ near the Fermi level $\epsilon=0$ for several values of $U$ at $T=0.001$ (a) 
and for several values of $T$ at $U=1.8$.}
\label{fig:r-dos}
%\vspace{+0.2cm}
\end{center}
\end{figure}

Figure \ref{fig:r-dos}(a) shows the renormalized $f$-DOS $\rho^{f}(\epsilon)$ near the Fermi level $\epsilon=0$ for several values of $U$ at a low temperature $T=0.001$. 
In this calculation, 
we perform the analytic continuation of the self-energy $\Sigma_m(z)$ from the imaginary frequency to the real frequency by using the Pad$\acute{\rm e}$ approximation, 
and then substitute $\Sigma_m(\epsilon+i0_{+})$ 
into Eq. (\ref{eq:gf}) to obtain $\rho^{f}(\epsilon)\equiv\rho_{m}^{f}(\epsilon)=-\frac{1}{\pi}{\rm Im}~G_{m}^{f}(\epsilon+i0_{+})$. 
When $U$ increases, the renormalized pseudogap with the renormalized characteristic energies $\tilde{{\it \Delta}}_1$ and $\tilde{{\it \Delta}}_2$, 
which correspond to ${\it \Delta}_1$ and ${\it \Delta}_2$ for $U=0$ shown in Fig. \ref{fig:b-dos}, 
decreases together with decrease in the quasiparticle band width resulting in the highly reduced pseudogap accompanied by the heavy-fermion bands for large $U$. 
We also find that the spectral weight due to the quasiparticle bands decreases with decreasing the quasiparticle band width, 
while that due to the broad peaks corresponding to the Hubbard-like bands around $\epsilon_{f}=-U/2$ and $\epsilon_{f}+U=U/2$ increases (not shown). 
We note that, as the self-energy $\Sigma_m(z)$ is independent of ${\bm k}$ in the DMFT, 
$\rho^{f}(0)$ at $T=0$ is unchanged by $U$ and has a finite value (see also Fig. \ref{fig:b-dos}) 
resulting in a metallic behavior at low temperature as mentioned in Sec. \ref{subsec:2B}.

One of the most remarkable features of the Kondo semiconductor is that the gap structure largely depends on temperature 
in contrast to the case with the ordinary semiconductor. 
In Fig. \ref{fig:r-dos}(b), we plot the renormalized $f$-DOS $\rho^{f}(\epsilon)$ near the Fermi level $\epsilon=0$ for several values of $T$ at $U=1.8$. 
At low temperature, we observe a clear pseudogap structure with the renormalized characteristic energies $\tilde{{\it \Delta}}_1$ and $\tilde{{\it \Delta}}_2$. 
With increasing $T$, the pseudogap structure is found to be smeared, and then finally disappears at high temperature above the so-called coherence temperature $T_0$, 
where the $T$ dependence of $\rho^{f}(\epsilon)$ is mainly caused by the evolution of the imaginary part of the self-energy ${\rm Im}\Sigma(\epsilon)$ 
that becomes large for $T \simg T_0$ as explicitly shown in Sec. \ref{subsec:3C}. 
Such $T$ dependence of $\rho^{f}(\epsilon)$ has been observed in the tunneling\cite{TS} and photoemission\cite{CeRhAs2,CeRhAs3} spectroscopies 
for the Kondo semiconductors.

\begin{figure}[t]
\begin{center}
\includegraphics[width=6.50cm]{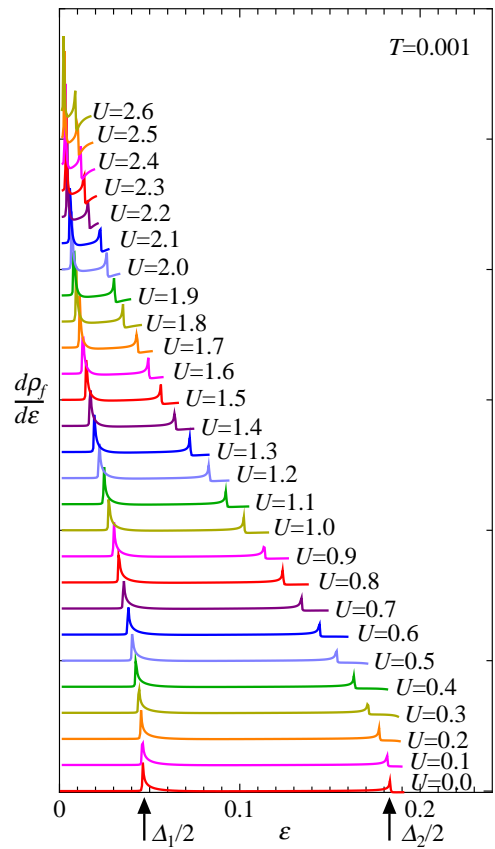}
\vspace{-0.5cm}
\caption{(Color online) The energy derivative of $f$-DOS $d\rho^{f}(\epsilon)/d\epsilon$ for several values of $U$.}
\label{fig:d-dos}
%\vspace{+0.2cm}
\end{center}
\end{figure}

\begin{figure}[t]
\begin{center}
\includegraphics[width=6.50cm]{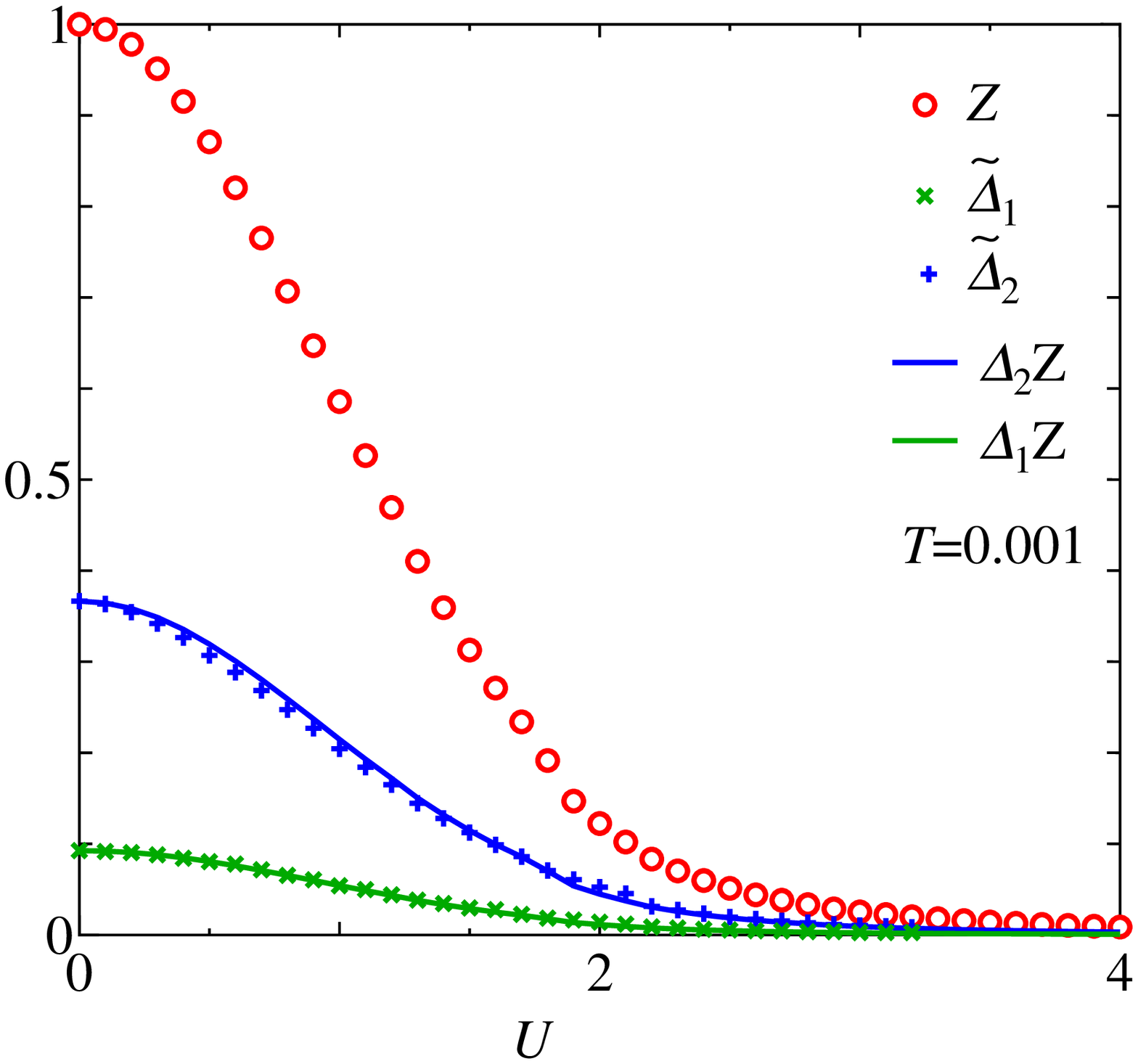}
\vspace{-0.5cm}
\caption{(Color online) $U$-dependence of the characteristic energies of the renormalized pseudogap $\tilde{{\it \Delta}}_1$ 
and $\tilde{{\it \Delta}}_2$ and the renormalization factor $Z$ together with ${\it \Delta}_1 Z$ and  ${\it \Delta}_1 Z$, 
where ${\it \Delta}_1$ and ${\it \Delta}_2$ are the bare characteristic energies of the pseudogap.}
\label{fig:gap}
%\vspace{+0.2cm}
\end{center}
\end{figure}

To see the renormalized characteristic energies $\tilde{{\it \Delta}}_1$ and $\tilde{{\it \Delta}}_2$ more explicitly, 
we plot the energy derivative of the $f$-DOS $d\rho^{f}(\epsilon)/d\epsilon$ for several values of $U$ at $T=0.001$ as shown in Fig. \ref{fig:d-dos}. 
The two significant peaks of $d\rho^{f}(\epsilon)/d\epsilon$ corresponding to $\tilde{{\it \Delta}}_1/2$ and $\tilde{{\it \Delta}}_2/2$ 
are clearly observed and found to decrease with increasing $U$. 
In Fig. \ref{fig:gap}, the $U$ dependence of $\tilde{{\it \Delta}}_1$ and $\tilde{{\it \Delta}}_2$ are plotted 
together with the renormalization factor $Z$ calculated from the real part of the self-energy 
as $Z\equiv Z_{m}=\left(1-\frac{d}{d\epsilon}{\rm Re}\Sigma_{m}(\epsilon)\Bigr|_{\epsilon=0}\right)^{-1}$. 
When $U$ increases, $\tilde{{\it \Delta}}_1$ and $\tilde{{\it \Delta}}_2$ decrease with decreasing $Z$ resulting in the highly reduced pseudogap. 

As mentioned in Sec. \ref{subsec:2B}, 
the pseudogap energies ${\it \Delta}_1$ and ${\it \Delta}_2$ are given by the band energies with $\hat{k}_z=0$ and $\hat{k}_z=\pm \sqrt{7/15}$, respectively. 
Then, $\tilde{{\it \Delta}}_1$ and $\tilde{{\it \Delta}}_2$ are expected to be given by the corresponding energies of the renormalized quasiparticle bands 
whose widths are reduced by the renormalization factor $Z$. 
In fact, $\tilde{{\it \Delta}}_1$ and $\tilde{{\it \Delta}}_2$ are found to be in good agreement with ${\it \Delta}_1 Z$ and ${\it \Delta}_2 Z$, respectively, 
and are highly reduced in proportion to $Z\ll 1$ for large $U$ as shown in Fig. \ref{fig:gap}. 
The highly reduced pseudogap observed for large $U$ is accompanied by the heavy fermions with the large mass enhancement factor $m^*/m=Z^{-1}\gg 1$ 
resulting in the heavy-fermion semiconductor with the nodal gap structure.

%This is because that $Z_{m}$ obtained by the DMFT is $\bm{k}$-independent, 
%and this is considrered to be widely satisfied in the heavy fermion systems.
%This is because the renormalization factor of each $\bm{k}$-quasiparticle is the same value as $Z$ in the DMFT 
%results in the same $U$-dependece between $Z$ and $\tilde{{\it \Delta}}_1~\tilde{{\it \Delta}}_2$.
%this means that the renormalization effect described $Z$ calculated by the present DMFT are properly introduced into the $f$-DOS throgh the $\bm{k}$-dependent $c$-$f$ mixing.
%the renormalization of $\tilde{{\it \Delta}}_1$ and $\tilde{{\it \Delta}}_2$  obey the $U$-dependence of $Z$ which have been understood as expected in the early study.\cite{ACF1}

%\if0
%Fig.2%%%%%%%%%%%%%%%%%%%%%%%%%%%%%%%%%%%%%%%%%%%%%%%%%%%%%%%%%%%%%%%%%%%%%%%%%%%%%%%
%%%%%%%%%%%%%%%%%%%%%%%%%%%%%%%%%%%%%%%%%%%%%%%%%%%%%%%%%%%%%%%%%%%%%%%%%%%%%%%%%%%%%
%Fig.3%%%%%%%%%%%%%%%%%%%%%%%%%%%%%%%%%%%%%%%%%%%%%%%%%%%%%%%%%%%%%%%%%%%%%%%%%%%%%%%
%%%%%%%%%%%%%%%%%%%%%%%%%%%%%%%%%%%%%%%%%%%%%%%%%%%%%%%%%%%%%%%%%%%%%%%%%%%%%%%%%%%%%
%Fig.4%%%%%%%%%%%%%%%%%%%%%%%%%%%%%%%%%%%%%%%%%%%%%%%%%%%%%%%%%%%%%%%%%%%%%%%%%%%%%%%
%%%%%%%%%%%%%%%%%%%%%%%%%%%%%%%%%%%%%%%%%%%%%%%%%%%%%%%%%%%%%%%%%%%%%%%%%%%%%%%%%%%%%
%\fi

\subsection{Magnetic and charge susceptibilities\label{subsec:3B}}
\begin{figure}[t]
\begin{center}
\includegraphics[width=6.75cm]{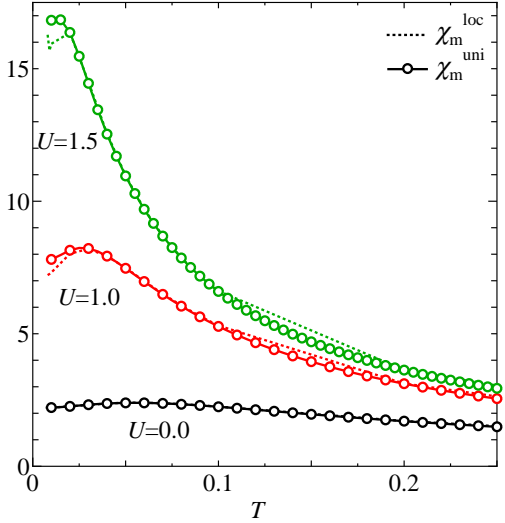}
\vspace{-0.5cm}
\caption{(Color online) $T$ dependence of the uniform and the local components of the magnetic susceptibilities 
for $f$ electrons $\chi_{\rm m}^{\rm uni}$ and $\chi_{\rm m}^{\rm loc}$ for $U=0$ and $U=1$.}
\label{fig:chiu}
%\vspace{+0.2cm}
\end{center}
\end{figure}

In Fig. \ref{fig:chiu}, we show the uniform and the local components of the magnetic susceptibilities for $f$ electrons 
$\chi_{\rm m}^{\rm uni}$ and $\chi_{\rm m}^{\rm loc}$ as functions of $T$ for $U=0,1$ and $1.5$, where $\chi_{\rm m}^{\rm uni}$ is calculated 
from the magnetization $M=\langle n_{+}^{f}\rangle-\langle n_{-}^{f}\rangle$ in the presence of a small external magnetic field $H=0.01$ 
as $\chi_{\rm m}^{\rm uni}=M/H$ and $\chi_{\rm m}^{\rm loc}$ is calculated from the eigenvalues and the eigenvectors in the effective impurity model for $H=0$ 
using the standard linear-response formulation. 
We can see both of the magnetic susceptibilities agree well with each other in the case with the present model where the specific nesting vector responsible for 
the strong $\bm{k}$-dependence of the magnetic susceptibility is absent. 

For $U=0$, $\chi_{\rm m}^{\rm uni}$ (together with $\chi_{\rm m}^{\rm loc}$) shows a maximum at a certain temperature $T_{\rm max}$ 
which roughly corresponds to ${\it \Delta}_1/2$ and is described by the Pauli paramagnetism with the pseudogap structure of the $f$-DOS (see Fig. \ref{fig:b-dos}). 
As shown in Fig. \ref{fig:chiu}, $\chi_{\rm m}^{\rm uni}$ ($\chi_{\rm m}^{\rm loc}$) is enhanced for $U=1$ and $U=1.5$ 
due to the correlation effect resulting in an enhanced Pauli paramagnetism at low temperature below the coherence temperature $T_0$, 
where the pseudogap structure of the renormalized $f$-DOS is observed [see Fig. \ref{fig:r-dos}(b)], 
and then, $\chi_{\rm m}^{\rm uni}$ ($\chi_{\rm m}^{\rm loc}$) shows a maximum at $T_{\rm max}\sim \tilde{\it \Delta}_1/2$ due to the pseudogap structure.
On the other hand, at high temperature $T \simg T_0$, 
$\chi_{\rm m}^{\rm uni}$ ($\chi_{\rm m}^{\rm loc}$) exhibits a Curie-law behavior $\chi_{\rm m}^{\rm uni}\sim\chi_{\rm m}^{\rm loc}\sim 1/T$ 
where the $f$ electron is considered to be almost localized 
and then the pseudogap structure in the renormalized $f$-DOS is found to disappear as shown in Fig. \ref{fig:r-dos}(b). 

\begin{figure}[t]
\begin{center}
\includegraphics[width=8.25cm]{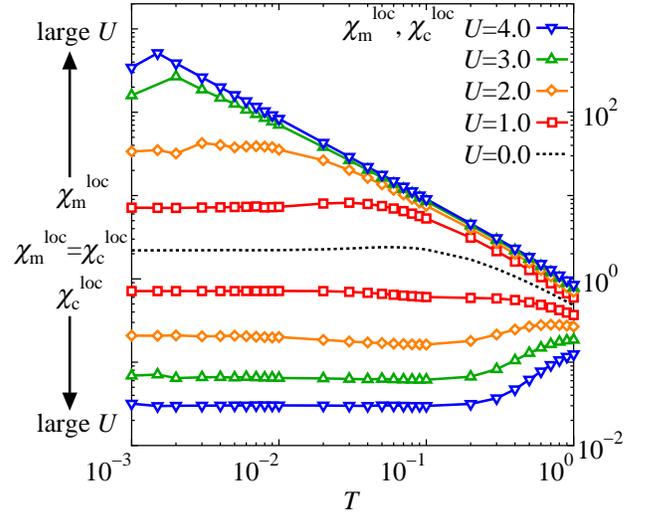}
\vspace{-0.5cm}
\caption{(Color online) $T$ dependence of the local magnetic and charge 
susceptibilities $\chi_{\rm m}^{\rm loc}(T)$ and $\chi_{\rm c}^{\rm loc}(T)$, respectively, for several values of $U$.}
\label{fig:chil}
%\vspace{+0.2cm}
\end{center}
\end{figure}

To see the correlation effect due to $U$ systematically, we plot the $T$ dependence of $\chi_{\rm m}^{\rm loc}$ for several values of $U$ in Fig. \ref{fig:chil}. 
When $U$ increases, $\chi_{\rm m}^{\rm loc}$ increases to show the enhanced Pauli paramagnetism at low temperature $T \siml T_0$ together 
with the Curie-law behavior at high temperature $T \simg T_0$, where $\tilde{\it \Delta}_1$ decreases with increasing $U$ as shown in Fig. \ref{fig:gap}. 
In Fig. \ref{fig:chil}, we also plot the $T$ dependence of the local charge susceptibility for $f$ electrons $\chi_{\rm c}^{\rm loc}$, 
which is also calculated from the eigenvalues and the eigenvectors in the effective impurity model using the standard linear-response formulation. 
$\chi_{\rm c}^{\rm loc}$ monotonically decreases with increasing $U$ and is largely suppressed for large $U$ corresponding to the Kondo regime. 

%\if0
%Fig.5%%%%%%%%%%%%%%%%%%%%%%%%%%%%%%%%%%%%%%%%%%%%%%%%%%%%%%%%%%%%%%%%%%%%%%%%%%%%%%%
%%%%%%%%%%%%%%%%%%%%%%%%%%%%%%%%%%%%%%%%%%%%%%%%%%%%%%%%%%%%%%%%%%%%%%%%%%%%%%%%%%%%%
%Fig.6%%%%%%%%%%%%%%%%%%%%%%%%%%%%%%%%%%%%%%%%%%%%%%%%%%%%%%%%%%%%%%%%%%%%%%%%%%%%%%%
%%%%%%%%%%%%%%%%%%%%%%%%%%%%%%%%%%%%%%%%%%%%%%%%%%%%%%%%%%%%%%%%%%%%%%%%%%%%%%%%%%%%%
%\fi

\subsection{Quasiparticle lifetime\label{subsec:3C}}
The quasiparticle lifetime $\tau$ is known to be related to the imaginary part of the self-energy as $1/2\tau=-{\rm Im}\Sigma(i0_{+})$. 
To obtain ${\rm Im}\Sigma(i0_{+})$, 
we perform the analytic continuation of the self-energy $\Sigma_m(z)$ from the imaginary frequency to the real frequency 
by using the Pad$\acute{\rm e}$ approximation as mentioned in Sec. \ref{subsec:3A}. 
It is found that $-{\rm Im}\Sigma(i0_{+})$ obeys the $T^2$ dependence at low temperature below the coherence temperature $T_0$ 
and is well reproduced by a fitting function $c_{0}+c\left(T/T_{0}\right)^{2}$, 
where $c_{0}$ the value at $T=0$ expected to be zero in the Fermi liquid theory and $c$ is the value at $T=T_{0}$. 

\begin{figure}[t]
\begin{center}
\includegraphics[width=8.5cm]{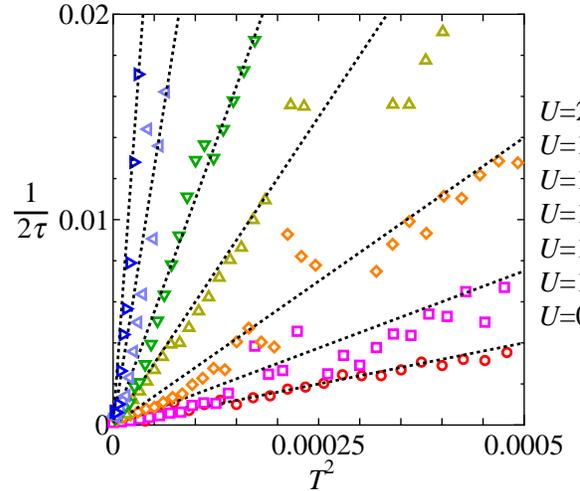}
%\vspace{-0.5cm}
%\vspace{+0.2cm}
\end{center}
\caption{(Color online) $T^2$-dependence of the inverse lifetime for several values of $U$. 
The dotted lines represent the fitting functions $1/2\tau=c\left(T/T_{0}\right)^{2}$ (dotted lines) with $c=0.04$ 
and the coherence temperature $T_{0}$ which depends on $U$.}
\label{fig:tau}
\end{figure}

In the present numerical calculation, $c_{0}$ is small ($c_{0}\ll c$) but finite due to the effect of finite size $N_s$, 
and is found to decrease with increasing $N_s$ as approaching $c_{0}\to 0$ for $N_s \to \infty$. 
Then, the inverse lifetime is estimated as $1/2\tau=-{\rm Im}\Sigma(i0_{+})-c_0$ and is plotted as a function of $T^2$ for several values of $U$ in Fig. \ref{fig:tau}. 
We can see that $1/2\tau$ thus obtained is in good agreement with the fitting function $1/2\tau=c\left(T/T_{0}\right)^{2}$ (dotted lines), 
where we set $c=0.04$ above which $1/2\tau$ is found to deviate from the 
$T^2$ dependence and we determine $T_0$ so as to fit $1/2\tau$ to the fitting function as possible for each $U$. 
When $U$ increases, the coherence temperature $T_0$ decreases in proportion to $Z$ as explicitly shown in the next section. 
%\if0
%Fig.7%%%%%%%%%%%%%%%%%%%%%%%%%%%%%%%%%%%%%%%%%%%%%%%%%%%%%%%%%%%%%%%%%%%%%%%%%%%%%%%
%%%%%%%%%%%%%%%%%%%%%%%%%%%%%%%%%%%%%%%%%%%%%%%%%%%%%%%%%%%%%%%%%%%%%%%%%%%%%%%%%%

\subsection{Characteristic temperatures\label{subsec:3D}}
\begin{figure}[t]
\begin{center}
\includegraphics[width=7.5cm]{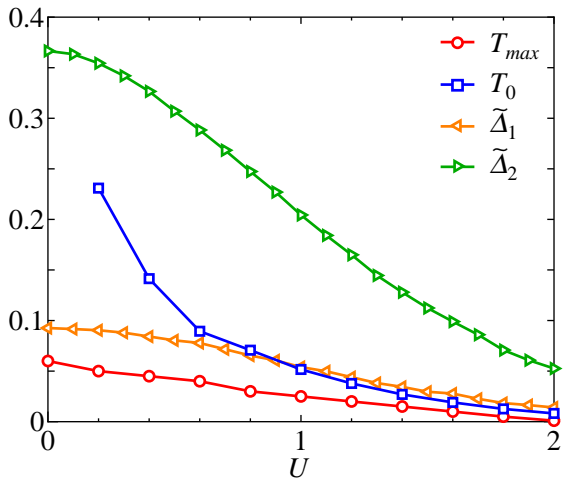}
\vspace{-0.5cm}
%\vspace{+0.2cm}
\end{center}
\caption{(Color online) $U$ dependence of the characteristic temperatures $T_{\rm max}$ and $T_{0}$ 
together with the characteristic energies of the renormalized pseudogap $\tilde{{\it \Delta}}_1$ and $\tilde{{\it \Delta}}_2$.}
\label{fig:ene}
\end{figure}

In Fig. \ref{fig:ene}, 
we plot the characteristic temperatures, $T_{\rm max}$ obtained in Sec. \ref{subsec:3B} and $T_{0}$ obtained in Sec. \ref{subsec:3C}, 
as functions of $U$ together with the characteristic energies of the renormalized pseudogap $\tilde{\it\Delta}_1$ and $\tilde{\it\Delta}_2$ 
obtained in Sec. \ref{subsec:3A}. 
We can see that $T_{0}$ monotonically decreases with increasing $U$ and is roughly proportional to $Z$ for large $U$, 
where $T_0 \sim \tilde{\it \Delta}_1 \sim {\it \Delta}_1 Z$ (see also Fig. \ref{fig:gap}). 
Then, the $T^2$ coefficient of the inverse lifetime $A=c/T_0^2$ is proportional to $Z^{-2}=(m^*/m)^2$ for large $U$ as expected from the Fermi liquid theory. 
We note that, for small $U$, the inverse lifetime, i. e., the imaginary part of the self-energy, can be obtained from the second-order perturbation 
with respect to $U$ and is proportional to $U^2$, and then $T_0 \propto U^{-1}$ as observed in Fig. \ref{fig:ene}. 

%\if0
%Fig.8%%%%%%%%%%%%%%%%%%%%%%%%%%%%%%%%%%%%%%%%%%%%%%%%%%%%%%%%%%%%%%%%%%%%%%%%%%%%%%%
%%%%%%%%%%%%%%%%%%%%%%%%%%%%%%%%%%%%%%%%%%%%%%%%%%%%%%%%%%%%%%%%%%%%%%%%%%%%%%%%%%%%%

At low temperature $T\siml T_0$, the imaginary part of the self-energy 
is sufficiently small to obtain the well defined quasiparticles, which yield the pseudogap structure of the renormalized $f$-DOS as shown in Fig. \ref{fig:r-dos} (b). 
Therefore the $T$ dependence of the physical quantities such as $\chi_{\rm m}$ is well described by the quasiparticle band for $T\siml T_0$, 
where $\chi_{\rm m}$ shows a maximum at $T_{\rm max}\sim \tilde{\it \Delta}_1/2$, which is smaller than $T_0$, 
due to the pseudogap structure of the renormalized $f$-DOS as shown in Fig. \ref{fig:chiu}. 
When $U$ increases, $T_{\rm max}\sim \tilde{\it \Delta}_1/2$ decreases in proportion to $Z$ as $\tilde{\it \Delta}_1 \sim {\it \Delta}_1 Z$ (see Fig. \ref{fig:gap}).

%%%
%\fi
\section{Results FOR $H\neq0$ \label{sec:4}}%%%%%%%%%%%%%%%%%%%%%%%%%%%%%%%%%%%%%%%%%%%%%%%%%%%%%%%
In this section, we examine the effect of the external magnetic field $H$ especially focused on the metamagnetic behavior. 
Due to the particle-hole symmetry with $\epsilon_{f}=-U/2$, the self-energy $\Sigma_m(z)$ is independent of $m$ even for $H \ne 0$, and then, $Z_{+}=Z_{-}\equiv Z$.

\subsection{Magnetization and differential susceptibility\label{subsec:4A}}
\begin{figure}[t]
\begin{center}
\includegraphics[width=8.75cm]{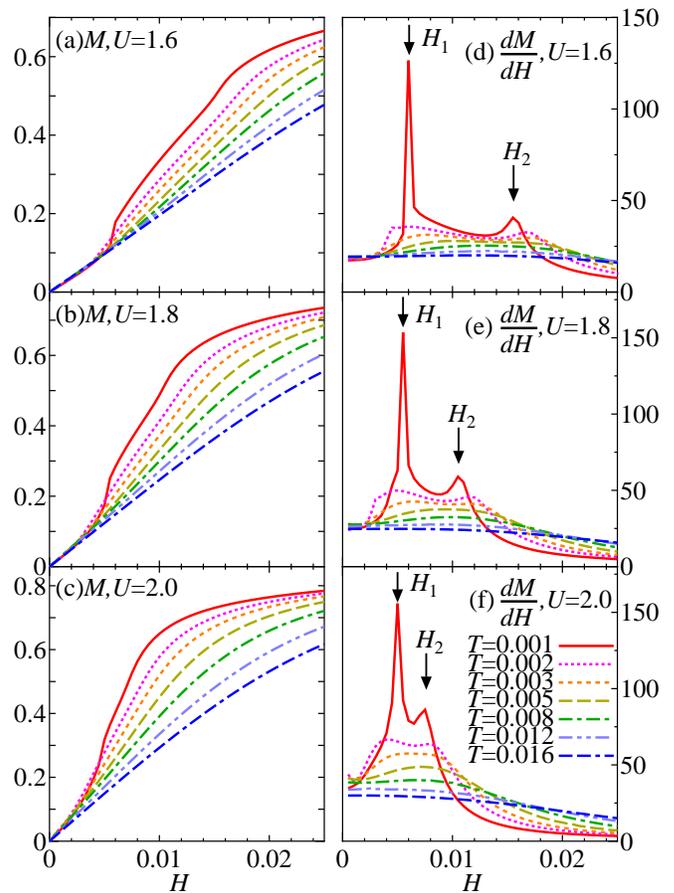}
\vspace{-0.5cm}
\caption{(Color online) $H$ dependence of the magnetization $M$ at $U=1.6$ (a), $U=1.8$ (b), and $U=2.0$ (c) 
and that of the differential susceptibility $dM/dH$  at $U=1.6$ (d), $U=1.8$ (e), and $U=2.0$ (f) for several values of $T$.}
\label{fig:mag}
%\vspace{+0.2cm}
\end{center}
\end{figure}

Figure. \ref{fig:mag} shows the $H$ dependence of the magnetization $M=\langle n_{+}^{f}\rangle-\langle n_{-}^{f}\rangle$ 
and that of the differential susceptibility $dM/dH$ for several values of $U$ and $T$.
At low temperature, two metamagnetic anomalies in $M$ are observed at 
critical magnetic fields $H=H_1$ and $H_2$ as shown in Figs. \ref{fig:mag}(a)$-$(c), 
and the corresponding two sharp peaks in $dM/dH$ are observed at $H=H_1$ and $H_2$ as shown in Figs. \ref{fig:mag}(d)$-$(f). 
We find that the critical magnetic fields $H_1$ and $H_2$ are given by the following relations: 
$H_1=\tilde{{\it \Delta}}_1(H_1)/2$ and $H_2=\tilde{{\it \Delta}}_2(H_2)/2$, respectively, 
where $\tilde{{\it \Delta}}_1(H)$ and $\tilde{{\it \Delta}}_2(H)$ are the renormalized pseudogap energies whose values for $H=0$ are shown in Fig. \ref{fig:gap} 
and largely depend on the external magnetic field $H$  as explicitly shown in the next section. 

When $U$ increases, $\tilde{{\it \Delta}}_1(H)$ and $\tilde{{\it \Delta}}_2(H)$ decrease for a given value of $H$ as explicitly shown in Fig. \ref{fig:gap} for $H=0$. 
Therefore $H_1$ and $H_2$ decrease with increasing $U$ as shown in Figs. \ref{fig:mag}(d)$-$(f). 
When $T$ increases, the two metamagnetic peaks in $dM/dH$ clearly observed at $T \siml T_{\rm max}$ are smeared, and then, 
merge into a broad peak at $T_{\rm max} \siml T \siml T_0$, where $T_{\rm max}$ and $T_0$ are shown in Fig. \ref{fig:ene}. 
The metamagnetic behavior finally disappears to show the monotonic $M$-$H$ curve similar to the case with the localized spins at high temperature $T\simg T_0$, 
where the pseudogap structure disappears as shown in Fig. \ref{fig:r-dos}(b) 
and the magnetic susceptibility shows the Curie-law behavior as shown in Fig. \ref{fig:chil}.

%\if0
%Fig.9%%%%%%%%%%%%%%%%%%%%%%%%%%%%%%%%%%%%%%%%%%%%%%%%%%%%%%%%%%%%%%%%%%%%%%%%%%%%%%%
%%%%%%%%%%%%%%%%%%%%%%%%%%%%%%%%%%%%%%%%%%%%%%%%%%%%%%%%%%%%%%%%%%%%%%%%%%%%%%%%%%%%
%\fi
\subsection{Inverse renormalization factor\label{subsec:4B}}
\begin{figure}[t]
\begin{center}
\includegraphics[width=7.5cm]{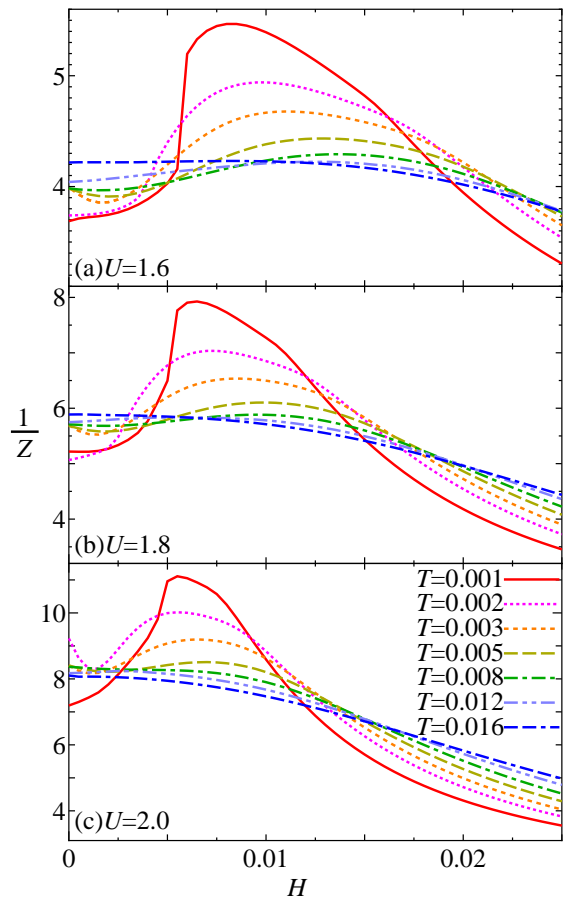}
\vspace{-0.5cm}
\caption{(Color online) $H$ dependence of the inverse renormalization factor $Z^{-1}$ for several values of $T$ at $U=1.6$ (a), $U=1.8$ (b) and $U=2.0$ (c).}
\label{fig:1/z}
%\vspace{+0.2cm}
\end{center}
\end{figure}

As mentioned in the previous subsection, the renormalized pseudogap energies $\tilde{{\it \Delta}}_1(H)$ and $\tilde{{\it \Delta}}_2(H)$ 
largely depend on the external magnetic field $H$ at low temperature. 
This is mainly caused by the $H$ dependence of the renormalization factor $Z(H)$ 
as $\tilde{{\it \Delta}}_1(H) \approx Z(H){\it \Delta}_1$ and $\tilde{{\it \Delta}}_2(H)\approx Z(H){\it \Delta}_2$. 
In Fig. \ref{fig:1/z}, we plot the $H$ dependence of the inverse renormalization factor $Z^{-1}$ for several values of $U$ and $T$. 
At low temperature, $Z^{-1}$ increases with increasing $H$ for $H\siml H_1$, where it shows an abrupt increase at $H=H_1$, 
and shows a peak for $H_1\siml H\siml H_2$, and then decreases for $H\simg H_2$, where it shows a kink at $H= H_2$. 
This $H$ dependence is caused by the pseudogap structure of the renormalized $f$-DOS $\rho^{f}(\epsilon)$ at low temperature 
(see Fig. \ref{fig:r-dos} for $H=0$), where the $f$-DOS at the chemical potential $\rho^{f}(0)$ is small due to the pseudogap for $H\siml H_1$, 
while it shows an abrupt increase at $H\sim H_1$ and becomes large for $H_1\siml H\siml H_2$, 
and then gradually decreases with increasing $H$ for $H\simg H_2$. 
Therefore the correlation effect between the $f$ electrons is enhanced due to $H$ especially for $H_1\siml H\siml H_2$ 
resulting in the enhancement of $m^*/m=Z^{-1}$. 
Then, we observe a remarkable field induced heavy-fermion state for $H_1\siml H\siml H_2$. 

When $T$ increases, $Z^{-1}$ decreases for $H_1\siml H\siml H_2$, while it increases for $H\siml H_1$ and $H\simg H_2$. 
Therefore, the peak structure of $Z^{-1}$ with pronounced anomalies at $H_1$ and $H_2$ clearly observed at low temperature $T\siml T_{\rm max}$ 
is smeared to show a broad peak around $H_1\siml H\siml H_2$ at $T_{\rm max} \siml T \siml T_0$, 
and then finally disappears to show the monotonically decreasing function of $H$ at high temperature $T\simg T_0$, as similar to the case with $dM/dH$ 
[see Figs. \ref{fig:mag}(d)$-$(f)]. 

\begin{figure}[t]
\begin{center}
\includegraphics[width=8.2cm]{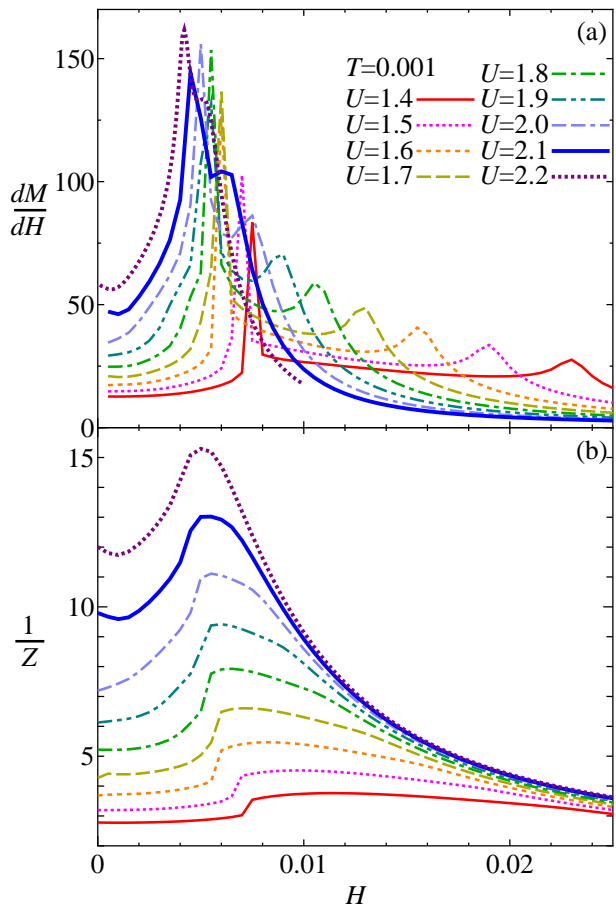}
\vspace{-0.5cm}
\caption{(Color online) $H$ dependence of the differential susceptibility $dM/dH$ (a) and inverse renormalization factor $Z^{-1}$ (b) 
for several values of $U$ at $T=0.001$.}
\label{fig:uhdep}
%\vspace{+0.2cm}
\end{center}
\end{figure}

Finally, we plot the differential susceptibility $dM/dH$ and the inverse renormalization factor $Z^{-1}$ 
as functions of $H$ for various $U$ at a low temperature $T=0.001$ in Figs. \ref{fig:uhdep}(a) and \ref{fig:uhdep}(b). 
At $H=H_1$, $dM/dH$ shows a sharp peak and $Z^{-1}$ shows an abrupt increase, while, at $H=H_2$, $dM/dH$ shows a cusp and $Z^{-1}$ shows a kink. 
For $H_1\siml H\siml H_2$, both of $dM/dH$ and $Z^{-1}$ are largely enhanced as compared to those values for $H=0$. 
Thus, we find that the pseudogap structure due to the point nodes of the hybridization gap 
in the PAM is responsible for the remarkable field induced heavy-fermion state accompanied by the metamagnetism. 

%\if0
%Fig.10%%%%%%%%%%%%%%%%%%%%%%%%%%%%%%%%%%%%%%%%%%%%%%%%%%%%%%%%%%%%%%%%%%%%%%%%%%%%%%
%%%%%%%%%%%%%%%%%%%%%%%%%%%%%%%%%%%%%%%%%%%%%%%%%%%%%%%%%%%%%%%%%%%%%%%%%%%%%%%%%%%%%
%Fig.11%%%%%%%%%%%%%%%%%%%%%%%%%%%%%%%%%%%%%%%%%%%%%%%%%%%%%%%%%%%%%%%%%%%%%%%%%%%%%%%
%%%%%%%%%%%%%%%%%%%%%%%%%%%%%%%%%%%%%%%%%%%%%%%%%%%%%%%%%%%%%%%%%%%%%%%%%%%%%%%%%%%%%
%\fi
\section{SUMMARY AND DISCUSSIONS\label{sec:5}}
%--------------------------------------summary---------------------------------------
In summary, we have investigated the PAM with the $\bm{k}$-dependent $c$-$f$ mixing 
that reproduces the point nodes of the hybridization gap simulating the pseudogap structure of the anisotropic Kondo semiconductors by using the DMFT+ED method. 
The physical quantities have been calculated systematically over the entire range of the parameters: $T$, $U$, and $H$. 
What we have found are as follows: 
(1) at low temperature below the coherence temperature $T_0$, the imaginary part of the self-energy is proportional to $T^2$, 
where the pseudogap with two characteristic energies $\tilde{\it \Delta}_1$ and $\tilde{\it \Delta}_2$ is observed. 
The magnetic susceptibility shows the enhanced Pauli paramagnetic behavior with a maximum at $T_{\rm max}(<T_0)$ due to the effect of the pseudogap. 
When $U$ increases, $\tilde{\it \Delta}_1$, $\tilde{\it \Delta}_2$, $T_0$ and $T_{\rm max}$ decrease in proportion to the renormalization factor $Z$ 
resulting in a heavy-fermion semiconductor with a large mass enhancement $m^*/m=Z^{-1}$ for large $U$. 
(2) In the presence of the external magnetic field $H$ at low temperature $T\siml T_0$, 
the magnetization $M$ shows two metamagnetic anomalies $H_1$ and $H_2$ corresponding to $\tilde{\it \Delta}_1$ and $\tilde{\it \Delta}_2$ 
that are reduced due to the effect of $H$ together with $Z$. Remarkably, $Z^{-1}$ is largely enhanced due to $H$ especially for $H_1 \siml H \siml H_2$, 
where the field induced heavy-fermion state is realized. 
(3) When $T$ increases, the pseudogap together with the metamagnetic anomalies is smeared 
due to the evolution of the imaginary self-energy and finally disappears at high temperature $T \simg T_0$, 
where the magnetic susceptibility shows the Curie-law behavior. 

%------------------------------about important findings and comparison with early studies------------------------------
The present DMFT results are consistent with the previous results from the Gutzwiller approximation\cite{Ikeda1996} 
and the slave-boson mean-field theory\cite{Moreno2000} at low temperature $T\ll T_0$ where the renormalized pseudogap is clearly observed. 
On the other hand, with increasing $T$, the deviation between the present 
and the previous results increases appreciably due to the evolution of the imaginary self-energy resulting in the smearing of the pseudogap, 
and then, the discrepancy becomes significant at high temperature $T\simg T_0$, 
where the pseudogap disappears in the present study in contrast to the previous studies with the rigid pseudogap structure. 

%The present calculation is focused only on the simplest case of the symmetric DOS with respect to the fermi level due to the half-filling electron number, which is the same condition in the earlier Gutzwiller study.\cite{Ikeda1996} The present method can be easily applied to the calculation away from the half-filling, where the asymmetric DOS is realized and we are also interested in such doping effect on the anisoropic Kondo semiconductor and would like to discuss else where.

In the previous DMFT studies, the PAM has been extensively investigated in the case with the $\bm{k}$-independent $c$-$f$ mixing,\cite{DMFT3,DMFT4,DMFT5} 
where the fully opened hybridization gap ${\it \Delta}$ is found to be highly reduced to a renormalized value $\tilde{\it \Delta}\approx Z {\it \Delta}$ 
due to the strong correlation effect. 
At the half-filling with $H=0$, the Fermi level sits in the renormalized hybridization gap resulting in the insulating ground state, 
i.e., the isotropic Kondo semiconductor. 
When $H$ increases, the DOS is split due to the Zeeman splitting, 
and then, the Fermi level enters the upper (lower) hybridized band of the majority (minority) spin above a critical magnetic field $H_c \sim \tilde{\it \Delta}/2$ 
at which the field induced insulator-metal transition takes place.\cite{DMFT4} 
Correspondingly, $Z^{-1}$ is found to be almost independent of $H$ below $H_c$, while $Z^{-1}$ shows a discontinuous increase at $H=H_{c}$,\cite{DMFT4} 
where the DOS at the Fermi level changes from zero to a finite value and then the correlation effect is enhanced. 
For $H>H_c$, $Z^{-1}$ monotonically decreases with increasing $H$ as the Kondo effect due to the local spin fluctuation is suppressed by the magnetic field. 
At finite temperature, 
the discontinuous increase in $Z^{-1}$ observed at $T=0$ changes into a broad peak around $H_{c}$ due to the thermal broadening effect.\cite{DMFT5}

In the present PAM with the $\bm{k}$-dependent $c$-$f$ mixing reproducing the anisotropic Kondo semiconductor, 
the Fermi level sits in the dip of the DOS at the half-filling instead of in the fully opened hybridization gap with the $\bm{k}$-independent $c$-$f$ mixing. 
Therefore the DOS at the Fermi level is finite even for $H=0$ and increases with increasing $H$ toward $H_1$. 
Correspondingly, the mass enhancement factor $Z^{-1}$ gradually increases with increasing $H$ toward $H_1$, 
where the correlation effect is enhanced due to the large DOS at the Fermi level. 
This is the origin of the field induced heavy-fermion state in the present model. 
For $H\gg H_1$, $Z^{-1}$ gradually decreases with increasing $H$ as the Kondo effect is suppressed by the magnetic field. 
We note that, in the present study, we concentrate only on the particle-hole symmetric case, where $Z_m$ is independent of $m$: $Z_{+}=Z_{-}\equiv Z$, 
even for finite $H$. In the case without the particle-hole symmetry, the $m$ dependence of $Z_m$ for finite $H$ is considered to be significant. 
In fact, 
such $m$ dependence of $Z_m$ was obtained in our previous DMFT study for the PAM away from the half-filling reproducing the heavy-fermion metamagnetism.\cite{TY1}

Finally, we compare our theoretical results with the experimental results in the anisotropic Kondo semiconductors such as CeNiSn. 
At low temperature $T\siml T_0$, 
the pseudogap structure due to the $\bm{k}$-dependent $c$-$f$ mixing well describe the characteristic $T$ dependence of the physical quantities observed in CeNiSn 
as previously obtained by Ikeda and Miyake.\cite{Ikeda1996} 
In addition, the present results account for the $T$ dependence also at high temperature $T\simg T_0$, e.g., 
the Curie law behavior of the magnetic susceptibility. 
Most significantly, the $T$ dependence of the $f$-DOS 
in the present results is consistent with the experimental results from the tunneling\cite{TS} and photoemission\cite{CeRhAs2,CeRhAs3} spectroscopies, 
where the pseudogap structure is clearly observed at low temperature, while it disappears at high temperature. 
The metamagnetic anomalies at $H_1$ and $H_2$ are also consistent with the experimental observations.\cite{MAG1,MAG2,MAG3} 
More recently, the high magnetic field measurements exceeding 50 T (see Ref. 22) have revealed 
that the existence of a third metamagnetic anomaly at $H_3(>H_2)$ which has not been obtained in the present study. 
To discuss the anomaly at $H_3$, the effect of the excited CEF levels, which are not included in the present study, is considered to be important. 
Therefore, we need further investigations to describe the $T$ and $H$ dependencies of the Kondo semiconductors 
by using more realistic models including the excited CEF levels together with the realistic band structure.

%------------------------------about AFM and CDW long range order------------------------------
%The AFM fluctuations\cite{NSCAT1,NSCAT2,NSCAT3} %and the pressure induced AFM\cite{CeNiSn-P} have been observed in this systems
%and recently, in the same class of compounds CeRhAs,\cite{CeRhAs1} the successive phase transitions\cite{CeRhAs2,CeRhAs3} also have been observed. 
%and the successive phase transitions\cite{CeRhAs2,CeRhAs3} have been reported in the same class of compounds. % as CeNiSn and CeRhAs. 
%Our method presented here is also available to study such long range orders together with the strong correlation effects and $\bm{k}$-dependent $c$-$f$ mixing. 
%In addition, the difference of the correlation effects on the metallic and insulating DOS 
%is an important topic and have been studied through the enhancement of the magnetic susceptibility and renormalization factor 
%for a few decades.\cite{Kontani1996,Kontani1997,Mutou1996} 
%We have also studied this effect within the same method at $T=0$, which will be presented in the subsequent paper.\cite{TY2}

%we need to improve the present model by taking into account the lattice effects.
%In the present DMFT study, we deal with $c$ electron as the plane wave for simplicity and, 
%the more precise lattice features can be included through the tight-binding like model.
%For the more profound understand to the Kondo semiconductor, the study of the effects is also highly desirable.

\begin{acknowledgments}
This work was partially supported by the Grant-in-Aid for Scientific
Research from the Ministry of Education, Culture, Sports, Science and
Technology of Japan. T. Y. is a research fellow of the Japan Society for
the Promotion of Science. 
\end{acknowledgments}

% Specify following sections are appendices. Use \appendix* if there only one appendix.
\appendix
\section{$\bm{k}$-dependent $c$-$f$ mixing matrix}
In this appendix, we note the general expressions of the $\bm{k}$-dependent $c$-$f$ mixing term.\cite{CF-MIX,Ikeda1996,Hanzawa2002,Miyazawa2003}
Here $c$ electron state is a direct product of the plane wave state and the spin state $\chi_{\sigma} (\xi)$ as
\begin{align}
&\psi_{\bm{k}\sigma}(\bm{r},\xi)=\langle  \bm{r}\xi|\bm{k}\sigma\rangle=\frac{1}{\sqrt{N}}e^{i\bm{k}\cdot\bm{r}}\chi_{\sigma} (\xi).
\end{align}
The plane wave expansion around a position vector $\bm{r}$ is given by
\begin{align}
e^{i\bm{k}\cdot\bm{r}}=4\pi\sum_{pq}i^{p}j_{p}(kr)Y_{pq}^{*}(\theta_{\bm{k}},\phi_{\bm{k}})Y_{pq}(\theta,\phi),
\end{align}
where $j_{p}(kr)$ is a spherical Bessel function and
$Y_{pq}(\theta,\phi)$ is a spherical harmonics with the argument of the
solid angle $\Omega_{\bm{r}}$ of the vector $\bm{r}$ or
$\Omega_{\bm{k}}$ of the wave vector $\bm{k}$. 
Therefore, the $c$ electron wave function is described by
\begin{align}
\psi_{\bm{k}\sigma}(\bm{r},\xi)=&\frac{4\pi}{\sqrt{N}}e^{i\bm{k}\cdot\bm{R}_i}\nonumber\\&\times\sum_{pq}i^{p}j_{p}(kr_{i})Y_{pq}^{*}(\Omega_{\bm{k}})Y_{pq}(\Omega_{\bm{r}_{i}})\chi_{\sigma} (\xi),
\end{align}
where $\bm{r}_{i}$ is a vector from a site $\bm{R}_{i}$, $\bm{r}_{i}=\bm{r}-\bm{R}_{i}$. 
%We prepare a rectangular DOS with a band width 2$D$ as the DOS of $c$ electron and an energy origin is set to 0,
%\begin{eqnarray}
%\rho(\epsilon)&=\frac{1}{N}\sum_{\bm{k}}\delta(\epsilon-\epsilon_{\bm{k}})=\frac{1}{2D}\theta(D-\epsilon)\theta(D+\epsilon)\nonumber\\
%&=\left\{ \begin{array}{ll}
%\frac{1}{2D} & |\epsilon|<D \\
%0 & |\epsilon|>D. \\
%\end{array} \right.
%\end{eqnarray}

The $f$ electron states are characterized by the localized atomic orbital around a site $\bm{R}_i$. 
The $LS$ coupling in the Ce compounds splits the 14-folded wave functions to the excited $J=7/2$ octet and the ground $J=5/2$ sextet. 
In Ce compounds, the sextet in $J=5/2$ are usually split into three Kramers doublets by CEF.
%In general, CEF splits $J$-multiplets $|JM\rangle$ into lower symmetry wave functions $|\Gamma\gamma\rangle$, where $\Gamma$ represents the
%irreducible representation of in a given point symmetry and $\gamma$ does the internal degree of freedom in the $\Gamma$, i.e. a different $\gamma$ state belonging to the same $\Gamma$ state is degenerate. 
Each Kramers doublet $|\mu\rangle$ is expanded with a linear combination of $J$ multiplets as below
%The wave function $|\Gamma\gamma\rangle$ in the CEF multiplets is expanded with a linear combination of $J$-multiplets as below
\begin{align}
%&\psi_{_{\Gamma\gamma}}(\bm{r},\xi)=\langle\bm{r}\xi|\Gamma\gamma\rangle=\sum_{M=-J}^{J}\langle\bm{r}\xi|JM\rangle\langle JM|\Gamma\gamma\rangle,\nonumber\\
%&=\sum_{Mq\sigma}b_{JM}^{\Gamma\gamma}a_{npq\sigma}^{JM}R_{np}(r)Y_{pq}(\theta,\phi)\chi_{\sigma} (\xi)\nonumber,
&\psi_{\mu}(\bm{r},\xi)=\langle\bm{r}\xi|\mu\rangle=\sum_{M=-J}^{J}\langle\bm{r}\xi|JM\rangle\langle JM|\mu\rangle,\nonumber\\
&=\sum_{Mq\sigma}b_{\mu M}a_{Mq\sigma}R_{np}(r)Y_{pq}(\theta,\phi)\chi_{\sigma}(\xi)\nonumber,
\end{align}
where $R_{np}(r)$ is the radial function with quantum number $(n,p)$
 and $a_{Mq\sigma}=\langle npq\sigma|JM\rangle$ is the Clebsch-Gordan coefficient
 and $b_{\mu M}=\langle JM|\mu\rangle$ is a coefficient with the linear combination of some of $|JM\rangle$.
%For simplicity, we express $a_{npq\sigma}^{JM}$ and $b_{JM}^{\Gamma\gamma}$ as $a_{Mq\sigma}$ and $b_{M\gamma}$ in the rest of the paper. 
Consequently, $f$ electron wave function around a site $\bm{R}_i$ is described by
\begin{align}
\psi_{\mu}(\bm{r}_i,\xi)
&=\sum_{Mq\sigma}b_{\mu M}a_{Mq\sigma}R_{np}(r_i)Y_{pq}(\Omega_{\bm{r}_i})\chi_{\sigma}(\xi),\nonumber
\end{align}
where $a_{Mq\sigma}=-\sigma\sqrt{\frac{7/2-\sigma M}{7}}\delta_{qM-\frac{\sigma}{2}}$ for the $J=5/2$ sextet.
%Let us derive the $\bm{k}$-dependent $c$-$f$ mixing $V_{\bm{k}m\sigma}$ in eq.(\ref{eq:Hcf}).
By using the above wave functions, $c$-$f$ mixing between the states
$|\bm{k}\sigma\rangle$ and $|\mu\rangle$ can be directly calculated as below,
\begin{align}
V_{\bm{k}\mu\sigma}^{i}&=\langle\bm{k}\sigma|v|\mu\rangle
=\frac{1}{\sqrt{N}}e^{-i\bm{k}\cdot\bm{R}_i}~V_{\bm{k}\mu\sigma},
\end{align}
where $V_{\bm{k}\mu\sigma}$ by
\begin{align}
V_{\bm{k}\mu\sigma}&=\sqrt{4\pi}V_{knp}\sum_{Mq}b_{\mu M}a_{Mq\sigma}Y_{pq}(\Omega_{\bm{k}}),\\
V_{knp}&=\sqrt{4\pi}(-i)^{p}\int dr_i~r_{i}^{2}j_{p}(kr_i)v(r_i)R_{np}(r_i),
\label{eq:Vcf}
\end{align}
where the mixing strength $V_{knp}=V_{cf}$ is a treated as a parameter of our model.
Thus, the $c$-$f$ mixing Hamiltonian between the states $|\bm{k}\sigma\rangle$ and $|\mu\rangle$ $H_{cf}$ 
is written by
\begin{align}
H_{cf}=\sum_{\bm{k}\mu\sigma}\left(V_{\bm{k}\mu\sigma}c_{\bm{k}\sigma}^{\dagger}f_{\bm{k}\mu}+{\rm H.c.}\right).
\end{align}

In general, the $J=5/2$ sextets are split into three Kramers doublets except for the $\Gamma_8$ state in the cubic CEF.
When we consider the only lowest Kramers doublets denoted by the pseudo spin states $\mu=\pm$, 
the $c$-$f$ mixing matrix $\hat{V}_{\bm{k}}$ with their elements 
$V_{\bm{k}\mu\sigma}$ scaled by $V_{cf}$ is a 2 $\times$ 2 matrix, which is written by
\begin{align}
\hat{V}_{\bm{k}}=\frac{1}{V_{cf}}\left(\begin{array}{cc}
V_{\bm{k}+\uparrow} &V_{\bm{k}+\downarrow}  \\
V_{\bm{k}-\uparrow} &V_{\bm{k}-\downarrow}  \\
\end{array}\right).
\end{align}
For example, when $\mu=\pm5/2$, 
\begin{align}
\hat{V}_{\bm{k}}=\left(\begin{array}{cc}
-\sqrt{\frac{4\pi}{7}}Y_{32}(\Omega_{\bm{k}})& \sqrt{\frac{24\pi}{7}}Y_{33}(\Omega_{\bm{k}}) \\
-\sqrt{\frac{24\pi}{7}}Y_{3-3}(\Omega_{\bm{k}})& \sqrt{\frac{4\pi}{7}}Y_{3-2}(\Omega_{\bm{k}}) \\
\end{array}\right),
\end{align}
and when $\mu=\pm3/2$,
\begin{align}
\hat{V}_{\bm{k}}=\left(\begin{array}{cc}
-\sqrt{\frac{8\pi}{7}}Y_{31}(\Omega_{\bm{k}})& \sqrt{\frac{20\pi}{7}}Y_{32}(\Omega_{\bm{k}}) \\
-\sqrt{\frac{20\pi}{7}}Y_{3-2}(\Omega_{\bm{k}})& \sqrt{\frac{8\pi}{7}}Y_{3-1}(\Omega_{\bm{k}}) \label{eq:Ik0}\\
\end{array}\right),
\end{align}
and when $\mu=\pm1/2$,
\begin{align}
\hat{V}_{\bm{k}}=\left(\begin{array}{cc}
-\sqrt{\frac{12\pi}{7}}Y_{31}(\Omega_{\bm{k}})& \sqrt{\frac{16\pi}{7}}Y_{30}(\Omega_{\bm{k}}) \\
-\sqrt{\frac{16\pi}{7}}Y_{30}(\Omega_{\bm{k}})& \sqrt{\frac{12\pi}{7}}Y_{3-1}(\Omega_{\bm{k}}) \\
\end{array}\right).
\end{align}
The important quantity for the $\bm{k}$-dependent $c$-$f$ mixing is $I_{\bm{k}}$ as mentioned in the Sec. \ref{subsec:2B}, 
and these are written by
\begin{align}
&\mu=\pm5/2~~I_{\bm{k}}=\frac{15}{8}V_{cf}^{2}(1-\hat{k}_{z}^{2})^{2},\\
&\mu=\pm3/2~~I_{\bm{k}}=\frac{3}{8}V_{cf}^{2}(1-\hat{k}_{z}^{2})(1+15\hat{k}_{z}^{2}),\label{eq:Ik}\\
&\mu=\pm1/2~~I_{\bm{k}}=\frac{3}{4}V_{cf}^{2}(1-2\hat{k}_{z}^{2}+5\hat{k}_{z}^{4}).
\end{align}

\bibliography{ks}

\end{document}